\documentclass[12pt,a4paper]{article}
\usepackage{amsmath,amsthm,amsfonts,amssymb,alltt}
\usepackage{graphics}
\usepackage{graphicx}
\usepackage{epsfig}
\newtheorem{theorem}{Theorem}[section]

\theoremstyle{definition}

\theoremstyle{remark}
\newtheorem{remark}[theorem]{Remark}

\numberwithin{equation}{section}

\begin{document}

\author{Vladimir Mityushev \\ \small{
Pedagogical University, ul. Podchorazych 2, Krakow 30-084, Poland} 
%\\
% tel. +48 12 6626339; fax: +48 12 6358858 
}
\title{Random composites and the generalized method of Schwarz I. Conductivity problems}
\date{}
\maketitle

\begin{abstract}
Two-phase composites with non-overlapping inclusions randomly embedded in matrix are investigated. A straight forward approach is applied to estimate the effective properties of random 2D composites. First, deterministic boundary value problems are solved for all locations of inclusions, i.e., for all events of the considered probabilistic space $\mathcal C$ by the generalized method of Schwarz. Second, the effective properties are calculated in analytical form and averaged over $\mathcal C$. This method is related to the classic method based on the average probabilistic values involving the $n$-point correlation functions. However, we avoid computation of the correlation functions and compute their weighted moments of high orders by an indirect method which does not address to the correlation functions. The effective properties are exactly expressed through these moments. It is proved that the generalized method of Schwarz converges for an arbitrary multiply connected doubly periodic domain and for an arbitrary contrast parameter. The proposed method yields effective in symbolic-numeric computations formulae of high order in concentration. In particular, the Torquato--Milton parameter $\zeta_1$ is exactly written for circular inclusions.  
\end{abstract}

\section{Introduction}\label{Intro}
The effective properties of random heterogeneous materials and methods of their computation are of considerable interest \cite{Adler}, \cite{andr1}, \cite{BKN}, \cite{andr2},  \cite{Kolpakov}, \cite{Milton}, \cite{Telega}, \cite{Torq}. Randomness in such problems reveals through random tensor--functions locally describing the physical properties of medium. Despite the considerable progress made in the theory of disordered media, the main tool for studying such systems remains numerical simulations. Frequently, it is just asserted that it is impossible to get general analytical formulae for the effective properties except dilute composites when interactions among inclusions are neglected and except regular composites having simple geometric structures. This opinion sustained by unlimited belief in numerics has to be questioned by the recent pure mathematical investigations devoted to explicit solution to the Riemann--Hilbert problem for multiply connected domains \cite{Mit1994}, \cite{Mit2011a}, \cite{MiR}  and by significant progress in symbolic computations \cite{CzNM2012}, \cite{CzNM2012-2}. In the present paper, we demonstrate that the theoretical results \cite{Mit1994}, \cite{Mit2011a}, \cite{MiR} can be effectively implemented in symbolic form that yields analytical formulae for random composites.  

For conductivity problems governed by Laplace's equation, the local conductivity tensor $\lambda(\mathbf x)$ can be considered as a random function of the spatial variable $\mathbf x=(x_1,x_2,x_3)$.
In the present paper, we restrict ourselves to two--phase composites with non--overlapping inclusions when a collection of particles with fixed shapes and sizes is embedded in matrix. More precisely, let be given a set of hard particles $\{\mathcal D_k, \;k=1,2,\ldots\}$ where each $\mathcal D_k$ has a fixed geometry. Let all the particles are randomly located in the space and each particle $\mathcal D_k$ occupies a domain $D_k$ without deformations. Thus, the deterministic elements $\mathcal D_k$ are introduced independently but the joint set $\{D_1,D_2,\ldots\}$ randomly. %Randomness arises in location of particles in a host material; therefore, it is strictly related to geometry. 
The diversity of random locations is expressed by joint probabilistic distributions of the non--overlapping domains $D_k$. 

Various methods were applied to such random composites.  The most known theoretical approach is based on the $n$--point correlation functions presented by Torquato \cite{Torq} and summarized below in Sec.\ref{SecClass}. This theory yields analytic formulae and bounds for the effective properties. The main shortage of this theory is computational difficulties to calculate the $n$--point correlation functions for large $n$. 

Let a distribution of inclusions be fixed. Then, the effective conductivity tensor $\Lambda_e$ is uniquely determined through $\mathcal D_k$ and their physical properties. In order to estimate $\Lambda_e$, various statistical approaches were applied to construct the representative volume element (RVE) \cite{Kanit}. These approaches are based on the straight forward computations of the effective properties for various locations of inclusions and  on the statistical investigations of the obtained numerical results. The main lack of this physical theory is that the notation of the RVE is not correctly defined and the results have not precise meaning. This theory is rather a statistical analysis of the computational experiments and physical measures. %Such a statistical approach require huge number of computational experiments. 
%
%This problem can be reduced to a two--dimensional problem for non--overlapping domains on the plane. Various methods were applied to calculate the local field and to estimate the macroscopic behaviour of circular cylinders  \cite{andr, andr1, andr2, Cheng, Crowdy, Fil, MRP}. 

The main purpose of this paper is to work out constructive analytical-numerical methods to calculate the effective constants of composites with non-overlapping inclusions and to develop the corresponding RVE theory. We apply the straight forward approach to estimate the effective conductivity of random 2D composites. First, deterministic boundary value problems are solved for all locations of inclusions, i.e., for all events in the considered probabilistic space $\mathcal C$ by the generalized method of Schwarz (GMS). This method was proposed by Mikhlin \cite{Mikhlin} as a generalization of the classical alternating method to multiply connected domains; convergence of the GMS was established in \cite{Mit1995}, \cite{MiR}. The GMS is referred to decomposition methods \cite{SBG} frequently used in numerical computations and realized in the form of alternating methods.  A method of integral equations closely related to the GMS was developed in \cite{BM2013}.

After solution to the deterministic problem by the GMS the ensemble average for the effective conductivity (mathematical expectation) is calculated in accordance with the given distribution. This method is related to the classic method \cite{Torq} based on the average probabilistic values involving the $n$--point correlation functions. In our approach following the GMS, we avoid computation of the correlation functions and compute their weighted moments. The effective properties are exactly written through these moments. Analytical formulae and numerical simulations demonstrate advantages of our approach in the case of non--overlapping inclusions. For instance for circular inclusions, our method yields effective in computations formulae of order $O(\nu^{20})$ in concentration $\nu$ (exactly written in Sec.\ref{Ex1} up to $O(\nu^8)$). Constructive formulae of the classic approach \cite{Torq} are deduced only up to $O(\nu^5)$ for not high contrast parameters. In particular, the Torquato--Milton parameter $\zeta_1$ \cite{Milton}, \cite{Torq} is exactly written for circular inclusions.  

Stochastic 2D problems are posed and solved in doubly periodic statement in the plane. Theoretically, doubly periodic problems constitute the special class of problems in the plane with infinite number of inclusions \cite{Mit1999}. However, the number of inclusion per a periodicity cell, $N$, is arbitrary taken, and the final formulae for the effective tensor contains $N$ in symbolic form. Similar non--periodic statements can be used following \cite{Mit2013b}. Such an approach explains, for instance, the divergent sum (integral) arising in applications of self consistent methods when the divergent sum does really diverge for some distributions. This implies that the corresponding composites cannot be homogenised over the whole plane even if the concentration is properly defined. The homogenization theory (see \cite{Telega} and works cited therein) justifies existence of the effective properties for statistically homogeneous random fields which constitute a subclass of heterogeneous fields discussed in \cite{Torq}, \cite{Mit1999}, \cite{Mit2013b}. %(see constructive formulae in \cite{Mit2013b} for non--overlapping inclusions). 
The divergence and other similar effects do not arise for doubly periodic composites when homogenization is always valid. A periodicity cell can be treated as a representative cell and vice versa. From practical point of view, each sample is finite, hence, it can be considered as a representative cell with many inclusions. Application of the new RVE theory (see the next paragraph) can reduce the number of inclusions per periodicity cell but not always significantly. Because reduction to a small number of inclusions per cell can distort the effective properties of the original sample. It was noted in \cite{BiM2001} that periodic arrange of inclusions decreases the effective conductivity when conductivity of inclusions is greater than conductivity of host.
Moreover, the theory of elliptic functions can be applied to constructive symbolic computations for periodic problems that yields exact and approximate analytical formulae for the effective tensor. 

The proposed method leads to construction of the rigorous RVE theory for non--overlapping inclusions \cite{Mit2006}. The effective tensor can be written in the form of expansion in the moments of the correlation functions which can be considered as "basic elements" depending only on locations of inclusions. The RVE is defined as the minimal size periodicity cell corresponding
to the set of basic elements calculated for the composite. A simple fast algorithm to determine the representative cell for a given composite is based on reduction of the number of inclusions per periodicity cell having the same basic elements as the given composite. It follows from simulations \cite{CzNM2012}, \cite{CzNM2012-2} that for the uniform non--overlapping distribution of disks in order to reach high accuracy in the effective conductivity one needs to solve the corresponding boundary value problems at least for 64 inclusions per cell repeated at least 1500 times. These technical parameters essentially exceed possibilities of the traditional statistical approach to the RVE. 

Here, we are restricted by composites with non--overlapping inclusions. Other types of composites and porous media, for instance when $\lambda(\mathbf x)$ obeys the Gaussian distribution, the correlation functions and reconstructions of media give excellent practical results (see \cite{Adler} and works cited therein). 

The goal of this paper is to show that the results obtained by the GMS \cite{BiM2001}, \cite{BiM2005}, \cite{Mit1995}--\cite{MRP}, \cite{Szcz} can be considered as a constructive approach to resolve some problems of the classic theory of random composites \cite{Torq} for non--overlapping inclusions. The GMS allows to overcome myths frequently met in the theory of random composites:

1. It is asserted that only application of the correlation functions is the proper method to treat random composites. Hence, restrictions in this field are related to computational difficulties arising in numerical computations of the correlation functions. As it is noted above, we obviate the need for the correlation functions calculating the mathematical expectation of the effective tensor after complete solution to the deterministic problems. 

2. It is thought that the contrast parameter expansions are valid only for sufficiently small contrast parameters. However, it was first justified in \cite{Mit1993} that such expansions for plane conductivity problems are valid for all possible values of the contrast parameters.

3. Self consistent methods for random composites are considered as a simple and constructive alternative to the method of correlation functions. However, applications of the GMS demonstrates essential limitations of these methods \cite{Mit2013b}. More precisely, the self consistent methods do not capture interactions among inclusions. Even solution to the $n$--particle problem in the infinite plane can be applied only to diluted distributions of the clusters consisting of $n$ inclusions.  
 
This paper is the first part of the work devoted to the GMS described in general form in Sec.\ref{SecGen}. The special attention is paid to conductivity problems. Integral equations corresponding to the GMS are described in Sec.\ref{SecInt}. These equations can be solved by different iterative schemes. The first scheme is based on the contrast expansions (the classic approach is in Sec.\ref{Contrast} and the GMS form in Sec.\ref{Contrast2}). The second method uses the cluster expansions (the classic approach in Sec.\ref{Cluster} and the GMS form is in Sec.\ref{Cluster2}). Applications of the GMS to elasticity problems will be described in a separate paper. 

\section{Deterministic and stochastic approaches}\label{Prob}
Let $\omega_1$ and $\omega_2$ be the fundamental pair of periods on the complex plane $\mathbb C$ such that $\omega_1>0$ and $\textrm{Im}\; \omega_2 > 0$ where $\textrm{Im}$ stands for the imaginary part. The fundamental parallelogram $Q$ is defined by the vertices $\pm \omega_1/2$ and $\pm \omega_2/2$. Without loss of generality the area of $Q$ can be normalized to one, hence, 
\begin{equation}
\omega_1\textrm{Im}\; \omega_2=1.
\label{area} 
\end{equation}
The points $m_1 \omega_1+ m_2\omega_2$ ($m_1,m_2 \in \mathbb Z$) generate a doubly periodic lattice $\mathcal Q$ (torus) where $\mathbb Z$ stands for the set of integer numbers. 

Consider $N$ non--overlapping simply connected domains $D_k$ in the cell $Q$ with Lyapuniv's boundaries $L_k$ %(see Figure 1) 
and the multiply connected domain $D=Q \backslash \cup_{k=1}^N(D_k \cup L_k)$, the complement of all the closures of $D_k$ to $Q$. Each curve $L_k$ leaves $D_k$ to the left.  Let a point $a_k$ is arbitrary fixed in $D_k$ ($k=1,2,\ldots,N$). 

We study conductivity of the doubly periodic composite when the host $D+m_1 \omega_1+ m_2\omega_2$ and the inclusions $D_k+m_1 \omega_1+ m_2\omega_2$ are occupied by conducting materials. Without loss of generality the conductivity of matrix can be normalized to one. The respective conductivity of inclusions is denoted by $\lambda$. %The contact between the components is assumed to be perfect. 
Introduce the local conductivity as the function
\begin{equation}
\lambda(z) = \left\{
\begin{array}{llll}
1,\quad z \in D,
\\
\lambda, \quad z \in D_k, \; k=1,2,\ldots,N.
\end{array}
\right.
\label{local}
\end{equation}  
where $z=x_1+ix_2$ denote a complex variable.
The potentials $u(z)$ and $u_k(z)$ are harmonic in $D$ and $D_k$
($k=1,2,\ldots,N$) and continuously differentiable in closures of the considered domains. 
The conjugation conditions are fulfilled on the contact interface
\begin{equation}
u=u_k,\quad \frac{\partial u}{\partial n}=\lambda\frac{\partial
u_{k}}{\partial n}\quad\text{on } L_k, \quad k=1,2,\ldots,N,
\label{2.1}
\end{equation}
where $\partial /\partial n$ denotes the outward normal derivative to $L_k$. The external field is modelled by the quasi--periodicity conditions 
\begin{equation}
u(z+\omega_1)=u(z)+\xi_1,\;u(z+\omega_2)=u(z)+\xi_2,  \label{2.2}
\end{equation}
where $\xi_{\ell}$ ($\ell = 1,2$) are constants. The external field applied to the considered doubly periodic composites is given by the vector $\mathbf E_0 = -(\xi_1,\xi_2)$. %Below, we frequently take the condition \eqref{2.2} with $\xi_1=\omega_1$ and $\xi_2=0$
%\begin{equation}
%u(z+\omega_1)=u(z)+\omega_1,\;u(z+\omega_2)=u(z).  \label{2.3}
%\end{equation}

The problem \eqref{2.1}--\eqref{2.2} can be stated in the probabilistic context with randomly located inclusions. %according to a probabilistic distribution. 
For definiteness, consider a set of the domains $D_k$ as a realization of the random locations of a collection of particles  $\mathcal D_k$ with fixed shapes and sizes. It  corresponds to location of hard particles in a vessel. Following the definition (3.4) from \cite{Torq} we introduce the doubly periodic specific probability density $P(\mathbf a)$ associated with finding a configuration of inclusions with position $\mathbf a :=(a_1, a_2, \ldots,a_N)$. The point $a_k$ determines the position of the particle $\mathcal D_k$ in the cell $Q$. The set of all the configurations is denoted by $\mathcal C$. Let $\textrm{d} a_k$ denote a small area element about the point $a_k$ and $\textrm{d} \mathbf a =\textrm{d} a_1 \textrm{d} a_2 \ldots \textrm{d} a_N$.  Then $P(\mathbf a) \textrm{d} \mathbf a$ is equal to the probability of finding $a_1$ in $\textrm{d} a_1$, $a_2$ in $\textrm{d} a_2$, $\ldots$, $a_N$ in $\textrm{d} a_N$. 

%Let $f(\mathbf x, \mathbf a)$ be a scalar field on the spacial variable $\mathbf x \in Q$ depending on the configuration $\mathbf a$. The spacial average over the cell $Q$ is defined as the integral
%\begin{equation}
%F(\mathbf a)=\langle f(\mathbf x, \mathbf a) \rangle = \int_{\mathcal A} %f(\mathbf x, \mathbf a) \textrm{d}\mathbf x.
%\label{aver1}
%\end{equation}
The ensemble average is introduced as the expectation in a probabilistic space
\begin{equation}
\widehat{f} = \int_{\mathcal C} f(\mathbf a) P(\mathbf a) \textrm{d}\mathbf a.
\label{aver1}
\end{equation} 
The effective conductivity tensor of deterministic composites 
\begin{equation}
\Lambda_e = \left(
\begin{array}{llll}
\lambda_{11} \quad \lambda_{12}
\\
\lambda_{21} \quad \lambda_{22}
\end{array}
\right)
\label{EfTen}
\end{equation}
can be found in terms of the solutions of two independent problems \eqref{2.1}--\eqref{2.2}. For instance, one can take $\xi_1=\omega_1$, $\xi_2=\textrm{Re}\;\omega_2$ and $\xi_1=0$, $\xi_2=\textrm{Im}\;\omega_2$. In the first case \cite{BiM2005},
\begin{equation}
\lambda_{11} = 1+ 2\rho \sum_{k=1}^N \int_{D_k} \frac{\partial u_k} {\partial x_1}\; \textrm{d}x_1 \textrm{d}x_2.
\label{e11}
\end{equation}
Here, $\rho$ denotes the contrast parameter introduced by Bergman
\begin{equation}
\rho = \frac{\lambda-1}{\lambda+1}.
\label{rho}
\end{equation}

The effective conductivity tensor $\Lambda_e$ of stochastic composites is determined by application of the operator \eqref{aver1} to the deterministic values \eqref{EfTen}.
Let the probabilistic distribution of the domains $D_k$ ($k=1,2,\ldots,N$) in $Q$ be such that the corresponding two--dimensional composite is isotropic in macroscale, i.e., the expected effective conductivity of the composite is expressed by a scalar. It can be calculated as follows
\begin{equation}
\lambda_e=\widehat{\lambda_{11}} = \int_{\mathcal C} \lambda_{11}(\mathbf a) P(\mathbf a)\textrm{d}\mathbf a,
\label{E11}
\end{equation}
where the deterministic value $\lambda_{11}(\mathbf a)$ is calculated by \eqref{e11}.

\section{Classic approach}\label{SecClass}
As it is stressed in Introduction we apply the straight forward approach to determine $\lambda_e$. First, the deterministic problem \eqref{2.1}--\eqref{2.2} is solved for every configuration $\mathbf a \in \mathcal C$ by the GMS. Further, the probabilistic average \eqref{E11} is calculated. In order to compare the results obtained by this method and by others we follow the classic approach to random composites described and summarized by Torquato \cite{Torq}. We follow the book \cite{Torq} and do not present formulae from earlier classic works (see references and historical notes therein and also in \cite{Milton}) in order to avoid multiple notations. 
We stress that only two--dimensional two--phase dispersed composites with one phase embedded in the connected matrix are considered in the framework of the general theory. 

In the book \cite{Torq}, the problem \eqref{2.1}--\eqref{2.2} is written in terms of the electric conduction by use of the local electric current $\mathbf J$ and the intensity field $\mathbf E$.
%\begin{equation}
%\nabla \cdot \mathbf J=0, \quad \nabla \times \mathbf E= \mathbf 0.
%\label{j1}
%\end{equation}
The potential $u$ satisfies equation 
\begin{equation}
\mathbf E =-\nabla u. 
\label{eq:j3}
\end{equation}
The linear constitutive relation is fulfilled in $D$ and $D_k$
\begin{equation}
\mathbf J(z) = \lambda(z) \mathbf E(z).
\label{j2}
\end{equation}
Torquato \cite{Torq} and others distinguish two main methods: cluster expansion and contrast expansion discussed below.

\subsection{Cluster expansions}\label{Cluster}
This method is presented as a correction of the single--inclusion approach by taking into account interactions between pairs of particle, triplets and so forth. It is worth noting that only infinite number of inclusions on plane gives a correct result \cite{Torq}, \cite{Mit2013b}. Study of the finite number of inclusions on plane yields the effective properties only of dilute composites as it is demonstrated in \cite{Mit2013b}.

The intensity field $\mathbf E$ is considered as a function of $z$ with the configuration parameter $\mathbf a$ and denoted by $\mathbf E(z;\mathbf a)$. Let the constant external field $\mathbf E_0$ is fixed. Torquato describes the following cluster expansion (see (19.8) from \cite{Torq})
\begin{equation}
\mathbf E(z;\mathbf a) = \mathbf E_0+ \sum_{k=1}^N \mathbf M_1(z;k) \cdot \mathbf E_0 + \sum_{k=1}^N \sum_{k_1 \neq k}^N \mathbf M_2(z; k, k_1) \cdot \mathbf E_0 + \cdots.
\label{j3}
\end{equation}
Here, $\mathbf M_1(z; k)$ is the single--inclusion operator which accounts for the first order interactions over and above $\mathbf E_0$ and inclusions periodically equivalent to the $k$-th inclusion, i.e., to $D_k+m_1 \omega_1+ m_2\omega_2$ for all integer $m_1$ and $m_2$. The operator $\mathbf M_1(z; k)$ can be derived by the following two methods. The first method corresponds to the Maxwell approach applied to dilute composites. It consists of two steps. At the first step, a boundary value problem for one inclusion $D_k$ in the infinite plane is solved and the single--inclusion operator $\mathbf K_1(z; k)$ in the plane is constructed (see for instance, formulae (17.4) and (19.6) from \cite{Torq}). Next, $\mathbf M_1(z; k)$ is constructed via $\mathbf K_1(z; k)$ by a periodicity operator. The periodicity transformation $\mathcal P:\mathbf K_1(z; k) \mapsto \mathbf M_1(z; k)$ can be easily performed since $\mathbf K_1(z; a_k)$ is expressed via the dipole tensor. The periodicity operator transforms rational functions to elliptic functions  \cite{akh}; for instance, $\mathcal P:z^{-1} \mapsto \zeta(z)$ and $\mathcal P:z^{-2} \mapsto \wp(z)$. The second method to construct $\mathbf M_1(z; k)$ is based on the same scheme realized in the torus topology \cite{Mit2005}.  

The operator $\mathbf K_1(z; k)$, hence $\mathbf M_1(z; k)$, is  written in closed form for disk, ellipse and many other domains (see \cite{Torq} and works cited therein). It can be constructed for any shape $D_k$ in terms of the conformal mapping of $D_k$ onto the unit disk. In general, $\mathbf K_p(z; k, k_1,\ldots, k_p)$ and $\mathbf M_p(z; k, k_1,\ldots, k_p)$ are the $p$-inclusion operator which accounts the $p$th order interactions. 

The effective conductivity can be found through the polarization field defined by (19.3)--(19.4) from \cite{Torq}
\begin{equation}
\mathbf P(z) = [\lambda(z)-1]\mathbf E(z)
= \left\{
\begin{array}{llll}
(\lambda-1)\mathbf E(z),\quad z \in D,
\\
\mathbf 0, \qquad z \in D_k, \; k=1,2,\ldots,N.
\end{array}
\right. 
\label{p1}
\end{equation}
Following the classic approach \cite{Torq} we introduce the "double" average operator over the unit cell $Q$ and over the configuration space $\mathcal C$
\begin{equation}
\langle g \rangle = \int_{Q} \int_{\mathcal C} g(x_1,x_2; \mathbf a) \textrm{d}\mathbf a\;\textrm{d}x_1 \textrm{d}x_2.
\label{averQ}
\end{equation}
%In the classic theory of random composites, equation $\langle g \rangle = \widehat{g}$ is assumed to be fulfilled by ergodicity arguments. In other words, the spacial average coincides with the probabilistic average.  
The effective tensor \eqref{EfTen} can be defined by equation
\begin{equation}
\langle \mathbf P \rangle =  (\Lambda_e-\mathbf I) \cdot \langle \mathbf E\rangle,
\label{averE}
\end{equation}
where $\mathbf I$ is the unit tensor. Having used \eqref{averE} Torquato (see (19.33) from \cite{Torq} and earlier investigations cited therein) deduced formula for macroscopically isotropic composites
\begin{equation}
\lambda_e =  1+\sum_{n=1}^{\infty} \frac{1}{n!}\int_{\mathcal C} W_n(\mathbf a)\; \textrm{d}\mathbf a,
\label{averW}
\end{equation}
where $W_n(\mathbf a)$ is a complicated functional of the $n$--inclusion cluster operators $\mathbf M_n(z; k, k_1,\ldots, k_n)$ and the probability density $P(\mathbf a)$. It is worth noting that $W_n(\mathbf a)$ is derived through $\mathbf K_n(z; k, k_1,\ldots, k_n)$ in \cite{Torq}. Actually, our periodic approach yields the same result by rearrangement of the terms in each $W_n(\mathbf a)$. 

Let $|D_k|$ denote the area of $D_k$. The concentration of inclusions is defined in the following way
\begin{equation}
\label{eq:conc}
\nu = \sum_{k=1}^N |D_k|.
\end{equation} 
It is noted in \cite{Torq} that \eqref{averW} yields the volume--fraction expansion (see (19.34) from \cite{Torq})
\begin{equation}
\lambda_e =  1+\sum_{n=1}^{\infty} B_n \nu^n,
\label{averV}
\end{equation}
where the coefficients $B_n$ are multidimensional integrals on $\mathbf M_p(z; k, k_1,\ldots, k_p)$ for $p=1,2,\ldots,n$. Direct applications of the correlation functions yield analytical formulae and numerical results for $B_n$  $n=2,3,4$ (see \cite{Torq} and a discussion in \cite{CzNM2012}).

\subsection{Contrast expansions}\label{Contrast}
Exact contrast expansions can be presented by formula (20.1) from \cite{Torq} which after some modifications becomes 
\begin{equation}
\lambda_e =  1+\sum_{n=1}^{\infty} C_n \rho^n,
\label{contr1}
\end{equation}
where $\rho$ is the contrast parameter \eqref{rho}. Many authors have been thinking that the series \eqref{contr1} converges only for sufficiently small $\rho$. Though convergence of the series for local fields had been proved in \cite{Mit1993}, \cite{Mit1994} for circular inclusions and in \cite{MiR} (see Sec.4.9.2) for general shapes  for all admissible $|\rho| \leq 1$. This of course implies the convergence of \eqref{contr1}.      

In order to deduce the general representation \eqref{contr1} and estimate the first coefficients $C_n$ ($n=1,2,3,4$) Torquato \cite{Torq} derives an integral equation on $\mathbf E(z)$ which can be written in the form (compare to equation (20.17) from \cite{Torq} for the cavity intensity field)
\begin{equation}
\mathbf E(z) = \mathbf E_0+ \int_{Q} \textrm{d}z' \; \mathbf G(z,z')\cdot [\lambda(z)-1]\mathbf E(z'), 
\label{contr2}
\end{equation}
where $\mathbf G(z,z')$ denotes the periodic Green's function. Substitution of \eqref{local} into \eqref{contr2} yields
\begin{equation}
\mathbf E(z) = \mathbf E_0+ \rho(\lambda+1)\sum_{k=1}^N\int_{D_k} \textrm{d}z' \; \mathbf G(z,z')\cdot \mathbf E(z'). 
\label{contr3}
\end{equation} 
The method of successive approximations applied to \eqref{contr3} yields a series  on the contrast parameter $\rho$ which can be shortly written in the form
\begin{equation}
\mathbf E(z)= \sum_{n=0}^\infty \rho^n  \mathbf H^n \cdot \mathbf E_0,
 \label{eq:intEq2}
\end{equation} 
where the operator $\mathbf H$ is defined by the right hand part of \eqref{contr3}. Use of \eqref{p1} yields an analogous series for $\mathbf P(z)$. The tensor $\Lambda_e$ satisfies equation \eqref{averE} which can be written in the form (20.37) from \cite{Torq}. The latter formula for macroscopically isotropic media becomes (20.57) from \cite{Torq}. In our designations it reads as follows 
\begin{equation}
\rho^2 \nu^2 (\lambda_e-1)^{-1}(\lambda_e+1)=\nu \rho -\sum_{n=3}^{\infty} A^{(1)}_n \rho^n.  
\label{contr4}
\end{equation} 
The $n$--point coefficients $A^{(1)}_n$ are exactly expressed in terms of the $p$--point correlation functions ($p=2,3,\ldots,n$) (see formulae (20.38)-(20.41) and (20.59)-(20.60) from \cite{Torq}). Formula \eqref{contr4} can be written in the form \eqref{contr1}. 

The coefficients $A^{(1)}_n$ were calculated in \cite{Torq} as sums of multiple integrals containing the correlation functions which are presented also through multiple integrals. Calculation of these integral is the main difficulty for numerical application of \eqref{contr4}. 

\section{The generalized method of Schwarz}\label{GMS}
\subsection{General}\label{SecGen}
The generalized method of Schwarz (GMS) was proposed by Mikhlin \cite{Mikhlin} as a generalization of the classical alternating method of Schwarz for finitely connected domains. The GMS is based on the decomposition of the considered domain with complex geometry onto simple domains and subsequent solution to boundary value problems for simple domains. In our case, we have $N$ simple domains, the cell $Q$ with one inclusion $D_k$ ($k=1,2,\ldots,N$). In each step of the algorithm the boundary conditions for the $k$th simple domain are corrected by influence of the $m$th simple domains ($m \neq k$) computed at the precedent steps. %The external field has to be properly taken into account in the above algorithm. 

We now shortly present the GMS, first for the conductivity problem \eqref{2.1}--\eqref{2.2} and further for general linear problems. The ideal contact condition \eqref{2.1} can be rewritten in the form
\begin{equation}
u_0=u_k-u^{ext},\quad \frac{\partial u_0}{\partial n}=\frac{\partial
(u_{k}-u^{ext})}{\partial n}+ \rho(\lambda+1) \frac{\partial
u_{k}}{\partial n} \quad\text{on } L_k, \quad k=1,2,\ldots,N,
\label{eq0-2.1}
\end{equation}
where $u=u_0+u^{ext}$ is the decomposition of the potential $u$ onto the regular periodic part $u_0$ and the external field $u^{ext}$ corresponding to the quasi periodicity conditions \eqref{2.2}. In the case of the space problem the external field $u^{ext}$ has a singularity at infinity.

Introduce the function 
$$
f(t)=(\lambda+1) 
\frac{\partial u_{k}}{\partial n}(t), \quad t\in L_k \;(k=1,2,\ldots,N)
$$ 
and the domains $D^+:=\cup_{k=1}^n D_k$, $D^-:=D$. It is convenient to represent the harmonic functions in different domains as one piece-wise harmonic function $u_0(z)$ introduced above in $D=D^-$ and equal to $u_k(z)-u^{ext}(z)$ in $D_k \cup L_k$ ($k=1,2,\ldots,N$). Then, \eqref{eq0-2.1} can be considered as the jump problem  
\begin{equation}
u_0^+=u_0^-,\quad \frac{\partial u_0^+}{\partial n}=\frac{\partial
u_0^-}{\partial n}+ \rho f(t) \quad\text{on } L=\cup_{k=1}^n L_k,
\label{eq0-2.1j}
\end{equation}
where $u_0^+$ and $u_0^-$ correspond to the limit values of $u(z)$ when $z$ tends to $t$ from $D^+$ and $D^-$, respectively. The jump problem \eqref{eq0-2.1j} has a unique solution expressed in terms of the simple layer potential $P$ for the curve $L$ in the cell $Q$ in the torus topology. We have
\begin{equation}
u_0(z) = \rho (Pf) (z), \quad z\in D^+ \cup D^-.
\label{eq:eq0-3}
\end{equation}
The operator $P$ is considered as an operator in an appropriate functional space \cite{Fil1970},  \cite{Fil1991},  \cite{Fil1994},  \cite{Kress}. In the classic statement, $Q$ should be replaced by the whole space. The operator $P$ is decomposed onto the simple layer potentials $P_k$ along curves $L_k$ ($k=1,2,\ldots,N$) as $P=\frac{1}{\lambda+1}\sum_{k=1}^N P_k$. Here, the multiplier $\frac{1}{\lambda+1}$ is introduced for short form of the equations below. Then \eqref{eq:eq0-3} implies that
\begin{equation} 
u_k(z) = \rho \sum_{m=1}^N \left(P_m \frac{\partial u_m^+}{\partial n} \right) (z)+ u^{ext}(z), \quad z \in D_k\; (k=1,2,\ldots,N), 
\label{eq:eq0-4a}
\end{equation}
and
\begin{equation} 
u(z) = \rho \sum_{m=1}^N \left(P_m \frac{\partial u_m^+}{\partial n} \right)(z)+ u^{ext}(z), \quad z\in D.
\label{eq:eq0-4}
\end{equation}
%where it is assumed that $u^{ext}(z)$ is harmonic in all $D_k$. 
Equations \eqref{eq:eq0-4a} can be considered as a system of integral equations on the potentials $u_k(z)$  in the inclusions $D_k$ $(k=1,2,\ldots,N)$. If it is solved, the potential $u(z)$ is calculated in $D$ by \eqref{eq:eq0-4}.

\begin{remark}
Equations \eqref{eq:eq0-4a} correspond to the GMS in Mikhlin's form \cite{Mikhlin} when the method of successive approximations can diverge for closely spaced inclusions. The following slight modification yields a convergent algorithm for any location of inclusions and every contrast parameter ($|\rho| \leq 1$)
\begin{equation}
\begin{array}{rr}
u_k(z) = \rho \sum_{m=1}^N \left[\left(P_m \frac{\partial u_m^+}{\partial n} \right) (z) - \left(P_m \frac{\partial u_m^+}{\partial n} \right) (w)\right]+ 
u^{ext}(z) + 
\\ 
\rho \sum_{m=1}^N \left(P_m \frac{\partial u_m^+}{\partial n} \right) (w), \quad z\in D_k \cup L_k  \quad (k=1,2,\ldots,N),
\end{array}
\label{eq:eq0-5}
\end{equation}
where $w$ is a fixed point in $D$ (see explanations in Sec.4 of \cite{Mit2011a}).
\end{remark}

Equations \eqref{eq:eq0-3}--\eqref{eq:eq0-5} are written for potentials. Their differentiation yields equations on the intensity field
\begin{equation}
\mathbf E_k(z) = \rho \sum_{m=1}^N (Q_m \mathbf E_m) (z)+ \mathbf E_0(z), \quad z\in D_k \cup L_k  \quad (k=1,2,\ldots,N),
\label{eq:eq0-6}
\end{equation}
where $\mathbf E_k(z)=\mathbf E(z)$ in $D_k$ and $Q_m$ are appropriate operators (similar to double layer potentials). Equation \eqref{eq:eq0-6} is similar to \eqref{contr2}.

This GMS for harmonic functions can be extended to general problems governed by linear equations describing electrical and heat conduction, flow in porous media, viscous flow, elasticity and coupled fields in $\mathbb R^d$ ($d=2,3$). 

Let tensor fields $\mathbf U(z)$ in $D$ and $\mathbf U_k(z)$ in $D_k$ satisfy the contact conditions
\begin{equation}
\mathbf T \cdot \mathbf U = \mathbf T_k \cdot \mathbf U_k, \quad \mbox{on}\; L_k  \quad (k=1,2,\ldots,N),
\label{eq:eq0-7}
\end{equation}
where $\mathbf T$ and $\mathbf T_k$ are boundary operators involving physical parameters of materials occupying the domains $D$ and $D_k$, respectively. Let $\mathbf T_k$ is presented in the form $\mathbf T_k = \mathbf T- \boldsymbol{\rho}_k$, where $\boldsymbol{\rho}_k$ denotes a contrast parameter tensor. Then \eqref{eq:eq0-7} becomes
\begin{equation}
\mathbf U = \mathbf U_k- \mathbf T^{-1} \cdot \boldsymbol{\rho}_k \cdot \mathbf U_k, \quad \mbox{on}\; L_k  \quad (k=1,2,\ldots,N),
\label{eq:eq0-8}
\end{equation} 
Let $\mathbf P_k$ denote the simple layer potential corresponding to the considered linear equation for the domains $D_k$ and $D_k^-$ separated by $L_k$. Application of $\sum_{k=1}^N \mathbf P_k$ to  \eqref{eq:eq0-8} yields
\begin{equation}
\mathbf U_k(z) = \sum_{m=1}^N (\mathbf P_m \cdot \mathbf T^{-1} \cdot \boldsymbol{\rho}_m \cdot \mathbf U_m) (z)+ \mathbf U^{ext}(z), \; z\in D_k \cup L_k  \; (k=1,2,\ldots,N).
\label{eq:eq0-9}
\end{equation}

There are two different methods to solve equations \eqref{eq:eq0-9}. The first method is based on the direct iterations and corresponds to the contrast expansion in $\rho$ (compare to Sec.\ref{Contrast}). The second method is based on implicit iterations applied to the same equations \eqref{eq:eq0-9} written in the form
\begin{equation}
\begin{array}{rr}
\mathbf U_k(z)-(\mathbf P_k \cdot \mathbf T^{-1} \cdot \boldsymbol{\rho}_k \cdot \mathbf U_k) (z) = \sum_{m\neq k}^N (\mathbf P_m \cdot \mathbf T^{-1} \cdot \boldsymbol{\rho}_m \cdot \mathbf U_m) (z)+ \mathbf U^{ext}(z), 
\\ 
z\in D_k \cup L_k  \quad (k=1,2,\ldots,N).
\end{array}
\label{eq:eq0-10}
\end{equation} 

One--inclusion problem for $D_k$ $(k=1,2,\ldots,N)$ is solved at each step of iterations. This scheme corresponds to the cluster expansions discussed in Sec.\ref{Cluster}. 

\begin{remark}
Equations \eqref{eq:eq0-3} and \eqref{eq:eq0-10} are not reduced to the classic integral equations constructed in the framework of the potentials theory \cite{Kress}. It is another type of equations.
\end{remark}

\begin{remark}
Equations \eqref{eq:eq0-3}--\eqref{eq:eq0-4} and \eqref{eq:eq0-9}-- \eqref{eq:eq0-10} can be derived in the limit cases of inclusions (soft and hard inclusions in terminology \cite{Movchan}) by introduction of fictive potentials \cite{MiR}.
\end{remark}

\subsection{Integral equations}\label{SecInt}
In the present section, we discuss the GMS for the double periodic problem \eqref{2.1}--\eqref{2.2}. Following \cite{MitOpt} and \cite{BiM2005} we introduce the complex potentials $\phi (z)$ and $\phi_k(z)$ in such a way that 
\begin{equation}
\label{eq:comp}
u\left( z\right) =\textrm{Re}\;\phi (z),\;u_k\left( z\right) =\frac 2{\lambda+1}\textrm{Re}\;\phi _k\left( z\right). 
\end{equation}
Two real equations 
\eqref{2.1} can be written as one $\mathbb R$--linear condition (see for details \cite{Mit1999}, \cite{MitOpt} and \cite{BiM2005})
\begin{equation}
\phi \left( t\right) =\phi _k\left( t\right) -\rho \;\overline{\phi _k\left(
t\right) },\;t \in L_k,\;k=1,2,\ldots,N,  \label{2.6}
\end{equation}
where the bar denotes the complex conjugation. 
The unknown functions $\phi \left( z\right)$ and $\phi _k\left( z\right) $ are analytic in $D$ and $D_k$, respectively, and continuously differentiable in the closures of the domains considered. It follows from \eqref{2.2} that the function $\phi (z)$ is quasi periodic \cite{Mit2005}
\begin{equation}
\phi(z+\omega_j)-\phi(z)=\xi_j+id_j\quad (j=1,2), 
\label{2.3j}
\end{equation}
where $i$ stands for the imaginary unit, the constants
%\begin{equation}
%\label{eq:xi}
%\xi=i(\xi_1 \overline{\omega_2}-\xi_2 \omega_1),
%\end{equation}
$\xi_1$ and $\xi_2$ are taken from \eqref{2.2}, $d_1$ and $d_2$ are undetermined  real constants which should be found during solution to the problem.

The following formula was proved in \cite{MiR}
\begin{equation}
\frac{d}{dt}[\phi_k(t)] = -\overline{n^2(t)} \;\overline{\phi'_k(t)} ,\;t \in L_k.
 \label{eq:B3}
\end{equation}
Here, the differential operator is understood as $\frac{d}{dt} =  t'_s\;\frac{d}{ds}$, where $t=t(s)$ is a parametric equation of $L_k$ and $\phi'_k(t)$ is the boundary value of the complex derivative $\frac{\partial}{\partial z} = \frac{\partial}{\partial x_1}- i \frac{\partial}{\partial x_2}$. 
Differentiation of \eqref{2.6} and application of \eqref{eq:B3} yields 
\begin{equation}
\psi \left( t\right) =\psi _k\left( t\right) +\rho \overline{n_k^2(t)} \;\overline{\psi _k\left(
t\right) },\;t \in L_k,\;k=1,2,\ldots,N,  \label{2.6psi}
\end{equation}
where $\psi(z)=\phi'(z)$, $\psi_k(z)=\phi'_k(z)$ and $n_k(t)$ stands for the unit normal outward vector to $L_k$  expressed in terms of the complex values. The function $\psi(z)$ is doubly periodic. The $\mathbb R$--linear problem \eqref{2.6psi} holds also in the limit cases for perfectly conducting inclusions ($\rho =1$) and for isolators ($\rho =-1$). 

The $\mathbb R$--linear problem \eqref{2.6}--\eqref{2.3j} can be reduced to a system of integral equation in the following way. The cell $Q$ in the torus topology is divided onto two domains $D^+=\cup_{k=1}^n D_k$  (not connected) and $D^-=D$ (multiply connected). Let $L = \cup_{k=1}^n L_k$ denote the boundary of $D^+$ and  $\mu(t)$ be a H\"{o}lder continuous function on $L$. Let $\zeta(z)$ denote the  Weierstrass $\zeta$--function for which \cite{akh}
\begin{equation}
\zeta(z+\omega_j)-\zeta(z)=\delta_j \quad (j=1,2),
\label{qp}
\end{equation}
where $\delta_j=2\zeta\left(\frac{\omega_j}2\right)$. Following \cite{weil},  \cite{MRP} introduce the Eisenstein function 
\begin{equation}
E_1(z)= \sum_{m_1,m_2 \in \mathbb Z} \frac{1}{z+m_1 \omega_1+m_2 \omega_2},
\label{eq:E1}
\end{equation}
where the Eisenstein summation is used \cite{weil}. The Weierstrass and Eisenstein functions are related by formula
$E_1(z)= \zeta(z) - S_2 z$, where $S_2=\frac{2}{\omega_1}\zeta\left(\frac{\omega_1}2\right)$. It follows from Legendre's identity $\delta_1\omega_2-\delta_2\omega_1=2\pi i$ (see \cite{akh},  \cite{weil}) and \eqref{qp} that the jumps of $E_1(z)$ have the form 
\begin{equation}
E_1(z+\omega_1)-E_1(z)=0, \quad E_1(z+\omega_2)-E_1(z)=-\frac{2\pi i}{\omega_1}.
\label{eq:qp}
\end{equation}

The Cauchy--type integral on torus was introduced with the $\zeta$--function in the kernel \cite{Chib}. We introduce it in the slight different form using the Eisenstein function 
\begin{equation}
\Phi(z)= \frac{1}{2\pi i} \int_L \mu(t)E_1(t-z) \textrm{d}t+C z+C_0. \quad z \in D^{\pm},
\label{Cachy}
\end{equation}
where $\mu(t)$ is a H\"{o}lder continuous function on $L$, $C$ and $C_0$ are complex constants. 
The function $\Phi(z)$ is analytic in $D^{\pm}$ and quasi--periodic, i.e., it has constants jumps per periodicity cell
\begin{equation}
\Phi(z+\omega_1)-\Phi(z)= C \omega_1, \quad \Phi(z+\omega_2)-\Phi(z)= \frac{1}{\omega_1} \int_L \mu(t) \textrm{d}t+C \omega_2.
\label{eq:qp1}
\end{equation}
%The quasi--periodic term $Cz+C_0$ with constants $C$ and $C_0$ is added to \eqref{Cachy} for convenience.
The following Sochocki (Sokhotski--Plemelj) formulae take place on torus \cite{Chib} 
\begin{equation}
\Phi^{\pm} (t)= \pm \frac{1}{2}\; \mu(t) +\frac{1}{2\pi i} \int_L \mu(\tau)E_1(\tau-t) \textrm{d}\tau, \quad t \in L,
\label{Soch}
\end{equation}
where $\Phi^{\pm} (t)$ denote the limit values of $\Phi(z)$ on $L$ when $z \in D^{\pm}$ tends to $t\in L$, respectively. 
Formulae \eqref{Soch} imply that the function $\Phi(z)$ satisfies the jump condition 
\begin{equation}
\Phi^+(t) - \Phi^-(t)= \mu(t), \; t \in L. \label{jump}
\end{equation}

Equation \eqref{jump} can be considered as a $\mathbb C$--problem problem in a class of quasi--periodic functions with the given jump $\mu(t)$. It follows from \cite{Chib} that the general solution of \eqref{jump} up to an additive arbitrary constant has the form \eqref{Cachy}.

Consider \eqref{2.6} as the problem \eqref{jump} with $\mu(t)=\rho \;\overline{\phi_k(t)}$ on $L$, $\Phi(z) = \phi_k(z)$ in $D_k \subset D^+$ and $\Phi(z) = \phi(z)$ in $D = D^-$. 
Then, \eqref{Cachy} can be written in $D^+$ as follows
\begin{equation}
\phi_k(z) = \rho \sum_{m=1}^N \frac{1}{2\pi i} \int_{L_m} \overline{\phi_m(t)}E_1(t-z) \textrm{d}t+Cz+C_0,
\;z \in D_k \;(k=1,2,\ldots, N).
 \label{intEq}
\end{equation}
\eqref{Cachy} in $D^-$ becomes
\begin{equation}
\phi(z) = \rho \sum_{m=1}^N \frac{1}{2\pi i} \int_{L_m} \overline{\phi_m(t)}E_1(t-z) \textrm{d}t+Cz+C_0,
\quad z \in D.
 \label{intEqpsi}
\end{equation}
Equations \eqref{intEq}--\eqref{intEqpsi} are deduced from the $\mathbb R$--linear problem \eqref{2.1}--\eqref{2.2}. Using the same arguments one can derive analogous equations from the problem \eqref{2.6psi} on the complex flux 
\begin{equation}
\psi_k(z) = -\rho \sum_{m=1}^N \frac{1}{2\pi i} \int_{L_m} \overline{n^2(t)} \;\overline{\psi_m(t)}E_1(t-z) \textrm{d}t+C,
\;z \in D_k \;(k=1,2,\ldots, N),
 \label{eq:B1}
\end{equation}
\begin{equation}
\psi(z) = -\rho \sum_{m=1}^N \frac{1}{2\pi i} \int_{L_m} \overline{n^2(t)} \;\overline{\psi_m(t)}E_1(t-z) \textrm{d}t+C,
\;z \in D.
 \label{eq:B2}
\end{equation}
It is also possible to obtain these equations by differentiation of \eqref{intEq}--\eqref{intEqpsi} using integration by parts and formula \eqref{eq:B3}.

The integral equations \eqref{eq:B1} are considered as an equation in the following Banach space. First, the space $\mathcal C^{(0,\alpha)}$ of functions H\"{o}lder continuous on $L$ is considered ($0<\alpha\leq 1$). This is a Banach space endowed with the norm
\begin{equation}
\label{eq:sup}
||\omega|| = \sup_{t \in L} |\omega(t)| + 
\sup_{t_{1,2} \in L\; t_1 \neq t_2} \frac{|\omega(t_1) -\omega(t_2)|}
{|t_1-t_2|^{\alpha}}.
\end{equation}
Consider the closed subspace $\mathcal A \subset \mathcal C^{(0,\alpha)}$ consisting of functions analytically continued into all $D_k$. Therefore, every element of $\mathcal A$ is a function analytic in the domain $\cup_{k=1}^N D_k$ and H\"{o}lder continuous in its closure. It is worth noting that the domain $\cup_{k=1}^N D_k$ is not connected and convergence in $\mathcal A$ implies the uniform convergence of functions in $\cup_{k=1}^N (D_k \cup L_k)$. 

Equations \eqref{eq:B1} are written in the domains $D_k$ and can be continued to the boundary by use of \eqref{Soch}
\begin{equation}
\begin{array}{rr}
\psi_k(t) = -\rho \sum_{m=1}^N \left[\frac{1}{2}\;\overline{n_m^2(t)}\;\overline{\psi_m(t)} + \frac{1}{2\pi i} \int_{L_m} \overline{n_m^2(t)}\;\overline{\psi_m(t)}E_1(t-z) \textrm{d}t \right]+C,
\\
t \in L_k \;(k=1,2,\ldots, N).
\end{array}
 \label{intEqB}
\end{equation} 
Therefore, the integral equations in the space $\mathcal A$ have the form \eqref{eq:B1}, \eqref{intEqB}. It is proved in Appendix that equations \eqref{eq:B1}, \eqref{intEqB} have a unique solution. Solution to these equations corresponds to the GMS which can be realized in different forms as it is noted in Sec.\ref{SecGen}. 

The problem \eqref{2.6}--\eqref{2.3j} for the complex potentials has a unique solution up to an arbitrary additive constant $C_0$. The constants $\xi_1+id_1$, $\xi_2+id_2$ and $C_0$ disappear in the problem \eqref{2.6psi} after differentiation. The structure of the general solution of \eqref{2.6psi} has the form $\psi(z) = \xi_1\psi^{(1)}(z)+\xi_2\psi^{(2)}(z)$ (see \cite{Mit2005}), where the real constants $\xi_1$ and $\xi_2$ correspond to the external field (jumps of $u(z)$ per a cell). Hence, the $\mathbb R$--linear problem \eqref{2.6psi} has two $\mathbb R$--linear independent solutions. From the other side, these two  independent solutions can be constructed by \eqref{eq:B1}--\eqref{eq:B2} taking, for instance $C=1$ and $C=i$. Therefore, the complex constant $C$  can be expressed through the real constants $\xi_1$ and $\xi_2$.
In order to calculate the effective properties of macroscopically isotropic composites it is sufficient to fix $\xi_1$, $\xi_2$ and find $C$. For simplicity, we put
\begin{equation} 
\xi_1 = \omega_1, \quad \xi_2=\textrm{Re} \;\omega_2.
 \label{eq:C9}
\end{equation} 
Then the external field corresponds to the complex potential $\phi^{(ext)}(z)=z$ and to the complex flux $\psi^{(ext)}(z)=1$. In the vector form, this external flux becomes $(-1,0)$.
It follows from the first relation \eqref{eq:qp1} that 
\begin{equation}
C=1+\frac{i d_1}{\omega_1}, 
 \label{eq:C10}
\end{equation} 
where $d_1= \textrm{Im}[\phi(z+\omega_1)-\phi(z)]$. Put $z=-\frac 12(\omega_1+\omega_2)$ in this relation and calculate
\begin{equation}
d_1=  \textrm{Im} \int_{-\frac 12(\omega_1+\omega_2)}^{\frac 12(\omega_1-\omega_2)} \psi(t) \textrm{d}t= -\int_{-\frac 12(\omega_1+ \textrm{R}\omega_2)}^{\frac 12(\omega_1-\textrm{Re}\omega_2)} \frac{\partial u}{\partial x_2}\left(x_1-\frac{i \textrm{Im}\; \omega_2}2 \right) \;\textrm{d}x_1.
 \label{eq:C11}
\end{equation} 
The integral from the left hand part of \eqref{eq:C11} expresses the total flux through the low side of the parallelogram $Q$ parallel to the $x_1$--axis. It must be equal to zero because of the periodicity and that the external flux $(-1,0)$ vanishes in the $x_2$--direction. Therefore, $d_1=0$, hence $C=1$ by \eqref{eq:C10}.

The effective conductivity tensor of deterministic composites can be calculated through the complex potentials in the inclusions (see \cite{Mit1999}, \cite{MitOpt} and \cite{BiM2005})
\begin{equation}
\lambda_{11}- i \lambda_{12}=1+2\rho\sum_{k=1}^N \int_{D_k}\psi_k(x_1+ix_2)\;\textrm{d}x_1\textrm{d}x_2,  
\label{2.8gen}
\end{equation}
where $\psi_k(z)$ ($k=1,2,\ldots, N$) satisfy \eqref{eq:B1}--\eqref{eq:B2}, \eqref{intEqB} with $C=1$.
The value \eqref{2.8gen} depends on the locations of inclusions, i.e., on $\mathbf a \in \mathcal C$. The effective conductivity of macroscopically isotropic random composites is defined through the operator \eqref{aver1} applied to \eqref{2.8gen}  
\begin{equation}
\lambda_e= \widehat{\lambda_{11}},  
\label{eq:2.8gen}
\end{equation}
where the macroscopic isotropy implies that $\widehat{\lambda_{12}}$ vanishes and $\widehat{\lambda_{11}}=\widehat{\lambda_{22}}$.

\subsection{Contrast expansions for the GMS}\label{Contrast2}
 It is proved in Appendix that the unique solution of \eqref{eq:B1}, \eqref{intEqB} can be found by a method of successive approximations uniformly convergent in $\cup_{k=1}^N (D_k \cup L_k)$. Let $\Psi(z) = \psi_k(z)$ for $z \in D_k \cup L_k$ and $A$ denote the operator from the right hand part of \eqref{intEq}, \eqref{intEqB}. Then, equations \eqref{intEq}, \eqref{intEqB} can be shortly written as the following equation in the space $\mathcal A$
\begin{equation}
\Psi= \rho A \Psi +1.
 \label{intEqA}
\end{equation}
Its unique solution is given by the series
\begin{equation}
\Psi= \sum_{n=0}^\infty \rho^n A^n 1.
 \label{eq:intEqA}
\end{equation} 

The method of successive approximations can be also presented in the form of  the following iterative scheme
\begin{equation}
\begin{array}{llrr}
\psi^{(0)}_k(z) =1, 
\\
\psi^{(p)}_k(z) = -\rho \sum_{m=1}^N \frac{1}{2\pi i} \int_{L_m} \overline{n_m^2(t)}\;\overline{\psi^{(p-1)}_m(t)}E_1(t-z) \textrm{d}t+1, \; p=1,2,\ldots ,
\\
\\
\;z \in D_k\quad (k=1,2,\ldots, N).
\end{array}
 \label{intEqS1}
\end{equation}
When the functions $\psi_k(z)$ ($k=1,2,\ldots, N$) are found, the effective conductivity is calculated by \eqref{2.8gen}.

In order to compare equations \eqref{contr3} and \eqref{intEq}, \eqref{intEqB} we introduce the complex intensity field $E(z)=E_1(z)-iE_2(z)$ isomorphic to the vector field $\mathbf E(z)=(E_1(z),E_2(z))$. Then, the vector equation \eqref{contr3} can be written as a complex equation on $E(z)$. For definiteness, we consider the complex potential $\psi_k(z)$ in $D_k$.
It follows from \eqref{eq:j3}, \eqref{eq:comp} and $\psi_k(z)=\phi'_k(z)$ that 
\begin{equation}
 \label{eq:comp2}
E(z)=\frac{\partial u_k}{\partial x_1}-i \frac{\partial u_k}{\partial x_2}=\frac{2}{\lambda+1} \psi_k(z).
\end{equation} 
Therefore, the complex equation on $E(z)$ in $D_k$ can be written as the equation \eqref{intEq} on $\psi_k(z)$. We do not directly prove that equations \eqref{contr3} and \eqref{intEqA} are the same. But it is easy to show that the forms of their solutions \eqref{eq:intEq2} and \eqref{eq:intEqA} coincide. Since the solutions $\mathbf E$ and $\Psi$ coincide (more precisely, isomorphic) for $\mathbf E_0=(-1,0)$, the power series \eqref{eq:intEq2} and \eqref{eq:intEqA} in $\rho$ can be considered as the presentations of the same function for sufficiently small $|\rho|$. Therefore, the coefficients of \eqref{eq:intEq2} and \eqref{eq:intEqA} coincide. In particular, this fact implies that substitution of \eqref{eq:intEqA} into \eqref{2.8gen} yields a series which coincides with the series \eqref{contr1} obtained by the classic approach. It follows from \cite{Mit1993}, \cite{Mit1994}, \cite{MiR} that the contrast expansion \eqref{eq:intEqA} converges for all admissible $|\rho|\leq 1$. %Other advantages of the GMS are discussed in Sec.\ref{Discussion}. 

\subsection{Cluster expansions for the GMS}\label{Cluster2}
The GMS can be presented by other iterative scheme corresponding to the classic cluster expansions summarized in Sec.\ref{Cluster}. First, we rewrite \eqref{intEq} in the form 
\begin{equation}
\begin{array}{llr}
\psi_k(z) +  \frac{\rho}{2\pi i} \int_{L_k} \overline{n_k^2(t)}\;\overline{\psi_k(t)}E_1(t-z) \textrm{d}t =
\\
\\
-\rho \sum_{m \neq k}  \frac{1}{2\pi i} \int_{L_m} \overline{n_m^2(t)}\;\overline{\psi_m(t)}E_1(t-z) \textrm{d}t+1, \;
z \in D_k\;(k=1,2,\ldots, n).
\end{array} 
 \label{intEq1}
\end{equation}

Following Sec.\ref{SecInt} for every fixed $m =1,2,\ldots, N$ introduce the Banach space $\mathcal A_m$ of 
functions  analytic in $D_m$ and H\"{o}lder continuous in $D_m \cup L_m$. Let $f_m(z) \in \mathcal A_m$.  For each fixed $k =1,2,\ldots, N$ introduce the operators
\begin{equation}
(P_{km} f_m)(z) =  \frac{1}{2\pi i} \int_{L_m} \overline{n_m^2(t)}\;\overline{f_m(t)}E_1(t-z) \textrm{d}t, \; z \in D_k\; (m =1,2,\ldots, n). 
 \label{eq:op1}
\end{equation}
%The $\zeta$--function of Weierstrass can be expanded into the series \cite{akh}
%\begin{equation}
%\zeta(z) =\frac{1}{z}-\frac{g_2 z^3}{2^2\cdot 3 \cdot 5}  -\frac{g_3 z^5}{2^2\cdot 5 \cdot 7} - \ldots. 
% \label{eq:zeta}
%\end{equation}
%Using \eqref{eq:zeta} we can present $P_{kk}$ in the form
%\begin{equation}
%(P_{kk} f_k)(z) =  \frac{1}{2\pi i} \int_{L_k} \frac{\overline{n_k^2(t)}\;\overline{f_k(t)}}{t-z}\; \textrm{d}t + \int_{L_k} R_k(t-z)\overline{f_k(t)}\; \textrm{d}t, \; z \in D_k,
% \label{eq:op2}
%\end{equation}
%where the second integral operator is compact in $\mathcal A_k$.  \eqref{eq:op2} determines the following structure of the operator $P_{kk}$ 
%\begin{equation}
%P_{kk}=  S_k + R_k,
% \label{eq:op4}
%\end{equation}
One can check that the operator $P_{kk}$ is singular and $P_{km}$ ($m\neq k$) are compact in  $\mathcal A_k$. Equations \eqref{intEq1} can be shortly written in the form
\begin{equation}
(I+\rho P_{kk})\psi_k = -\rho \sum_{m \neq k} P_{km} \psi_m+1,\quad k =1,2,\ldots, n.
 \label{eq:intEq10}
\end{equation}
The zero-th approximation in the concentration $\nu$ for \eqref{intEq1} can be written as the following integral equation %on $\psi_k(z)$ 
\begin{equation}
\psi^{(0)}_k(z) + \frac{\rho}{2\pi i} \int_{L_k} \overline{n_k^2(t)}\;\overline{\psi^{(0)}_k(t)}E_1(t-z) \textrm{d}t =1, \quad z \in D_k.
 \label{intEq2}
\end{equation}
Equation \eqref{intEq2} corresponds to the $\mathbb R$--linear problem \eqref{2.6psi} for one inclusion $D_k$ in the cell $Q$. Solution to the one--inclusion problem \eqref{intEq2} defines the inverse operator $(I+\rho P_{kk})^{-1}$ bounded in $\mathcal A_k$. Then, \eqref{eq:intEq10} can be written in the equivalent form
\begin{equation}
\psi_k = -\rho \sum_{m \neq k} (I+\rho P_{kk})^{-1}P_{km} \psi_m+(I+\rho P_{kk})^{-1}1,\; k =1,2,\ldots, n.
 \label{eq:intEq11}
\end{equation}
As it is established in Sec.\ref{Contrast2} the vector--function $\mathbf E(z)$ and the complex functions $\psi_k(z)$ express the same vector field in $D_k$ up to an isomorphism for $\mathbf E_0=(-1,0)$. Therefore, application of the successive approximations to \eqref{eq:intEq11} yields the same series as \eqref{j3} in $D_k$. A detailed comparison can be performed. It shows, for instance, that the term $\mathbf E_0+ \sum_{k=1}^N \mathbf M_1(z;k) \cdot \mathbf E_0$ from \eqref{j3} corresponds to $(I+\rho P_{kk})^{-1}1$ from \eqref{eq:intEq11}.

The solution of the problem \eqref{intEq2} depends on the shape of $D_k$. Therefore, the first order approximation in $\nu$ for the effective conductivity also depends on the shape and can be calculated by \eqref{2.8gen}. If all the inclusions have the same form, the functions $\psi^{(0)}_k(z)$ do not depend on $k$. Then, we arrive at the formula valid up to $O(\nu^2)$
\begin{equation}
 \label{CMA1}
\lambda_e \approx 1+2\rho \nu\alpha,
\end{equation}
where the shape factor $\alpha$ is introduced following \cite{LL}, \cite{Mit2013b}
$$
\alpha =\frac{1}{|D_1|} \int_{D_1} \psi^{(0)}_1(x_1+ix_2) \textrm{d}x_1\textrm{d}x_2.
$$ 
Application of the Pad\'{e} $(1,1)$--approximation in $\nu$ to \eqref{CMA1} yields
\begin{equation}
 \label{CMAs}
\lambda_e \approx \frac{1+ \rho \nu \alpha}{1-\rho \nu \alpha}.
\end{equation} 
One can find limitations of the formulae \eqref{CMA1}, \eqref{CMAs} and a comparison of their accuracy in \cite{Mit2013b}. 

The higher order iterations in the cluster expansion are based on the integral equations on $\psi^{(p)}_k(z)$
\begin{equation}
\begin{array}{llr}
\psi^{(p)}_k(z) +  \frac{\rho}{2\pi i} \int_{L_k} \overline{n_k^2(t)}\;\overline{\psi^{(p)}_k(t)}E_1(t-z) \textrm{d}t =
\\
\\
-\rho \sum_{m \neq k}  \frac{1}{2\pi i} \int_{L_m} \overline{n_m^2(t)}\;\overline{\psi^{(p-1)}_m(t)}E_1(t-z) \textrm{d}t+1, 
\;
z \in D_k \;(k=1,2,\ldots, n); 
\\
\\
\quad p=1,2,\ldots.
\end{array} 
 \label{intEqP}
\end{equation}
The uniform convergence of the iterative scheme \eqref{intEqP} was proved 
in the case $\rho=1$ (perfectly conducting inclusions) in Sec.4.9.2 of \cite{MiR}. 
The higher order iterations \eqref{intEqP} yield approximate analytical formulae for $\lambda_e$ for arbitrary non--overlapping locations of different disks \cite{MitOpt}, \cite{BiM2001}, \cite{Szcz} \cite{BiM2005} and for ellipses \cite{Mit2009}. An example of equal disks is discussed below in Sec.\ref{Ex1}.

\subsection{Circular inclusions}\label{Ex1}
Let the inclusions $D_k$ be disks of radius $r$ centered at the points $a_k$. Then, the normal unit vector $n_k(t)$ to the circle $|t-a_k|=r$ has the form $n_k(t)=\frac{t-a_k}r$. The inversion with respect to $|t-a_k|=r$ is defined by formula 
\begin{equation}
 \label{eq:ex1}
z_{(k)}^*=\frac{r^2}{\overline{z-a_k}}+a_k.
\end{equation}     
Consider the integral operator from \eqref{intEq}
\begin{equation}
 \label{eq:ex2}
(J_{km}f)(z)=\frac{1}{2\pi i} \int_{L_m} \overline{n_m^2(t)}\;\overline{f(t)}E_1(t-z) \textrm{d}t, \quad z\in D_k,
\end{equation}
where $f(z)$ is analytic in the disk $D_m$. This operator can be written in the form
\begin{equation}
 \label{eq:ex3}
(J_{km}f)(z)=\frac{1}{2\pi i} \int_{L_m} \left(\frac{r}{t-a_m}\right)^2\;\overline{f\left(t_{(m)}^*\right)}E_1(t-z) \textrm{d}t, \quad z\in D_k,
\end{equation}
where the function $\overline{f\left(t_{(m)}^*\right)}$ is analytically continued into $|z-a_k|>r$. %Let the Taylor series of $f(z)$ in $D_k$ has the form 
%\begin{equation}
% \label{eq:ex6}
% f(z) = \sum_{n=0}^\infty f_n (z-a_m)^n.
% \end{equation}  
%Then $\overline{f\left(z_{(m)}^*\right)}$ is expanded into Taylor series  
%\begin{equation}
% \label{eq:ex7}
% \overline{f\left(z_{(m)}^*\right)} = \sum_{n=0}^\infty %\frac{\overline{f_n}}{(z-a_m)^n}, \quad |z-a_k|>r.
% \end{equation}  

%The Weierstrass $\zeta$--function is represented in the form 
%\begin{equation}
% \label{eq:ex4}
%\zeta(z) = \frac1z+ \sum_{m_1,m_2 \in \mathbb Z}\nolimits' \left(\frac1{z-m_1\omega_1 -m_2\omega_2} + \frac1{m_1\omega_1 +m_2\omega_2} + \frac{z}{(m_1\omega_1 +m_2\omega_2)^2} \right),
%\end{equation}
%where $m_1,m_2$ run over the integers except the term $(0,0)$. Hence, 
The function $E_1(z)$ has poles of first order at the points $z=m_1\omega_1 +m_2\omega_2$ (see \cite{weil}). The function $\left(\frac{r}{z-a_m}\right)^2\;\overline{f\left(z_{(m)}^*\right)}$ is analytic in the domain $|z-a_m|>r$ and the integral $(J_{km}f)(z)$ can be calculated by residues at $|z-a_m|>r$. Introduce the operator 
\begin{equation}
 \label{eq:ex5}
\left(W^{(m)}_{m_1,m_2}f \right)(z)=\left(\frac{r}{z+m_1\omega_1 +m_2\omega_2} \right)^2 \overline{f\left(a_m+ \frac{r^2}{\overline{z-a_m+m_1\omega_1 +m_2\omega_2}}\right)}
\end{equation}
We have for $m=k$ 
\begin{equation}
 \label{eq:ex5a}
(J_{kk}f)(z)=-\sum_{m_1,m_2 \in \mathbb Z}\nolimits' \left(W^{(k)}_{m_1,m_2}f \right)(z)
\end{equation}    
and for $m\neq k$
\begin{equation}
 \label{eq:ex8}
(J_{km}f)(z)=-\sum_{m_1,m_2 \in \mathbb Z}\left(W^{(m)}_{m_1,m_2}f \right)(z).
\end{equation}    
The double series \eqref{eq:ex5}--\eqref{eq:ex8} are defined by the Eisenstein summation that correctly determines their convergence \cite{BiM2001}, \cite{MRP}.

Substitution of \eqref{eq:ex5}--\eqref{eq:ex8} into \eqref{intEq} yields the system of functional equations
\begin{equation}
\psi_k(z) = \rho\sum_{m=1}^N \sum_{m_1,m_2 \in \mathbb Z}\nolimits^* \left(W^{(m)}_{m_1,m_2}\psi_m\right)(z)+1,
\;|z -a_k| \leq r\;(k=1,2,\ldots, N),
 \label{eq:ex9}
\end{equation}
where $m_1,m_2$ run over integers in the sum $\sum^*$ with the excluded term $m_1=m_2=0$ for $m=k$. The functional equations \eqref{eq:ex9} do not contain any integral. Each term $\left(W^{(m)}_{m_1,m_2}\psi_m\right)(z)$ can be treated as the shift operator written in the form of functional compositions. Such equations can be easily solved by use of symbolic computations. The difficulty related to the infinite double summations in $m_1,m_2$ can be overcome by application of the Eisenstein functions $E_n(z)$. Equations \eqref{eq:ex9} can be solved by the contrast expansions in $\rho$ (see Sec.\ref{Contrast2}) and by the cluster expansions in $\nu$ (see Sec.\ref{Cluster2}). We refer to \cite{MitOpt}, \cite{BiM2001}, \cite{BiM2005}, \cite{MRP} where this approach is explained in details. The computational efficiency of the method can be demonstrated by few analytical formulae below. Consider the cluster expansion in $\nu=N \pi r^2$ which is equivalent to the expansion in $r^2$ \cite{BiM2001}, \cite{BiM2005}
\begin{equation}
\psi_m(z) = \sum_{n=0}^\infty \psi_{m}^{(n)}(z) r^{2n},
\;|z -a_k| \leq r\;(k=1,2,\ldots, N).
 \label{eq:ex10}
\end{equation}
Few first functions in the expansion \eqref{eq:ex10} are given by the following exact formulae
\begin{equation}
\begin{array}{lll}
\psi_{m}^{(0)}(z) =1, \quad \psi_{m}^{(1)}(z) =\rho \sum_{k=1}^N E_2(z-a_k), 
\\ 
\\
\psi_{m}^{(2)}(z) =\rho^2 \sum_{k,k_1=1}^N \overline{E_2(a_k-a_{k_1})}E_2(z-a_k),
\\ 
\\
\psi_{m}^{(3)}(z) =\rho^3 \sum_{k,k_1,k_2=1}^N E_2(a_k-a_{k_1})\overline{E_2(a_{k_1}-a_{k_2})}E_2(z-a_{k_2})-
\\
\\
2\rho^2 \sum_{k,k_1=1}^N \overline{E_3(a_k-a_{k_1})}E_3(z-a_k),
\end{array}
 \label{eq:ex11}
\end{equation}
where $E_n(z-a_k)$ is defined as $\widetilde{E}_n(z-a_k)=E_n(z-a_k)-(z-a_k)^{-p}$ for $k=m$ and $E_n(0):=\widetilde{E}_n(0)$.
It is possible to proceed \eqref{eq:ex11} and calculate the next approximations following the algorithm \cite{BiM2001}, \cite{BiM2005}.

Let $q$ be a natural number; $k_{s}$ runs over $1$ to $N$, $m_q=2,3,\ldots$. Let $\mathbf{C}$ denote the operator of complex conjugation. Introduce the multiple convolution sums
\begin{equation}
e_{m_{1}...m_{q}}=\frac{1}{N^{[1+\frac 12(m_{1}+\cdots+m_{q})]}}
\sum_{k_{0}k_{1}...k_{q}}E_{m_{1}}(a_{k_{0}}-a_{k_{1}})\overline{
E_{m_{2}}(a_{k_{1}}-a_{k_{2}})}...\mathbf{C}
^{q}E_{m_{q}}(a_{k_{q-1}}-a_{k_{q}}).  \label{2.10}
\end{equation} 
The integrals from \eqref{2.8gen} are simplified by the mean value theorem for harmonic functions
\begin{equation}
\lambda_{11}- i \lambda_{12}=1+2\rho \nu  \frac{1}{N}\sum_{k=1}^N \psi_k(a_k).  
\label{eq:ex12}
\end{equation}
Substitution of \eqref{eq:ex11} and the next terms into \eqref{eq:ex12} yields \cite{Szcz}, \cite{MRP} 
\begin{equation}
\lambda_{11}- i \lambda_{12}=1+2\rho \nu (1+A_1\nu+A_2\nu^2+\ldots),  
\label{eq:ex13}
\end{equation}
where
\begin{eqnarray}
A_{1} &=&\frac{\rho }{\pi }e_{2},\quad
A_{2}=\frac{\rho ^{2}}{\pi ^{2}}%
e_{22},\; A_{3}=\frac{1}{\pi ^{3}}\left[ - 2\rho
^{2}e_{33}+\rho ^{3}e_{222}%
\right] ,\quad  \notag \\
A_{4} &=&\frac{1}{\pi ^{4}}\left[ 3\rho ^{2}e_{44}- 2\rho
^{3}(e_{332}+e_{233})+\rho ^{4}e_{2222}\right],\quad
 \notag
\\
A_5 &=& \frac{1}{\pi^5} \left[- 4\rho^2 e_{55}+
 \rho^3(3e_{442}+6e_{343}+3e_{244})-\right.
\notag \\
&&\left.- 2\rho ^{4}(e_{3322}+e_{2332}+e_{2233})+\rho
^{5}e_{22222}\right], \label{5,18}
\\
A_{6} &=&\frac{1}{\pi ^{6}}\left[ 5\rho ^{2}e_{66}- 4\rho
^{3}(e_{255}+3e_{354}+3e_{453}+e_{552})+ \right. \notag
\\ && +\rho
^{4}(3e_{2244}
+6e_{2343}+4e_{3333}+3e_{2442}+6e_{3432}+3e_{4422})-
\notag \\
&&\left.-2\rho ^{5}(e_{22233}+e_{22332}+e_{23322}+e_{33222})+\rho
^{6}e_{222222})\right]. \notag
\end{eqnarray}
The next coefficients $A_n$ can be written in closed form by application of the algorithm presented in \cite{Szcz}, \cite{MRP}. The deterministic values \eqref{2.10} provide the benchmark to compute the effective conductivity of random composites. Following the Monte Carlo method one can take a representative sets of $\mathbf a \in \mathcal C$ to statistically calculate the expectations
\begin{equation}
\widehat{e}_{m_{1}...m_{q}} = \int_{\mathcal C} e_{m_{1}...m_{q}}(\mathbf a) P(\mathbf a) \textrm{d}\mathbf a.
\label{eq:aver1}
\end{equation} 
The infinite set $\{\widehat{e}_{m_{1}...m_{q}}, m_{j}=2,3,\ldots\}$ completely determine the random geometric structure of the considered class of composites and can be taken as the basic set in the RVE theory \cite{Mit2006}.

A fast algorithm to compute the convolution sums \eqref{2.10} is described in \cite{CzNM2012-2}.  It allows to compute $A_n$ ($n \leq 20$) within a given distribution of $a_k$ in few hours on an ordinary notebook when $N=64$ and with the number of computational experiments $1500$ in the Monte Carlo method. It follows from \cite{CzNM2012} that six terms \eqref{5,18} are sufficient to deduce an analytic formula for a uniform non--overlapping distribution of the disks.

Formulae \eqref{eq:ex13}--\eqref{5,18} corresponds to the cluster expansion. Application of the explicit iterations to the functional equations \eqref{eq:ex9} yields the contrast expansion as a series in $\rho$. One can see that the following third order formula in $\rho$ takes place
%\begin{equation}
%\psi_{m}(z) =1+\rho r^2 \sum_{k=1}^N E_2(z-a_k)+O(\rho^2).
%\label{eq:ex13}
%\end{equation} 
%Substitution of \eqref{eq:ex13} into \eqref{eq:ex12} yields
\begin{equation}
\lambda_{11}- i \lambda_{12}=1+2\rho \nu  +2\rho^2 \nu^2 \frac{e_2}{\pi}+ 2\rho^3 \nu^3 \sum_{n=2}^\infty (-1)^n (n-1)e_{nn} \frac{\nu^{n-2}}{\pi^n}+ O(\rho^4). 
\label{eq:ex14}
\end{equation}
For macroscopically isotropic composites, the expectation of \eqref{eq:ex14} has the form
\begin{equation}
\lambda_e=1+2\rho \nu  +2\rho^2 \nu^2 + 2\rho^3 \nu^3 \sum_{n=2}^\infty (-1)^n (n-1)\widehat{e}_{nn} \frac{\nu^{n-2}}{\pi^n}+ O(\rho^4), 
\label{eq:ex15}
\end{equation} 
where the relation $\widehat{e}_2=\pi$ from \cite{Mit2012} is used. Another formula deduced in \cite{Mit2012} can be useful in simulations and estimations of the third order term of \eqref{eq:ex14}--\eqref{eq:ex15}
\begin{equation}
e_{nn}=\frac{(-1)^n}{N^{n+1}} \sum_{m=1}^N \left|\sum_{k=1}^N E_n(a_m-a_k) \right|^2.
\label{eq:ex16}
\end{equation} 
One can find the numerical values of $\widehat{e}_{nn}$ for uniform non--overlapping distributions in \cite{CzNM2012-2}.

In order to compare the contrast expansion formula \eqref{eq:ex15} and formula \eqref{contr4} of the classic theory, we write \eqref{contr4} up to $O(\rho^4)$
\begin{equation}
\rho^2 \nu^2 (\lambda_e-1)^{-1}(\lambda_e+1)=\nu \rho - A^{(1)}_3 \rho^3+O(\rho^4).  
\label{eq:ex17}
\end{equation} 
Substitution of \eqref{eq:ex15} into \eqref{eq:ex17} yields 
\begin{equation}
A^{(1)}_3 =\nu^3 \left( \sum_{n=2}^\infty (-1)^n (n-1)\widehat{e}_{nn}  \frac{\nu^{n-2}}{\pi^n}-1\right).  
\label{eq:ex18}
\end{equation}
The latter formula gives the Torquato--Milton parameter $\zeta_1$ (see (20.59) and (20.66) from \cite{Torq})
\begin{equation}
\zeta_1 =\frac{\nu^2}{1-\nu} \left[ \sum_{n=2}^\infty (-1)^n (n-1)\widehat{e}_{nn}  \frac{\nu^{n-2}}{\pi^n}-1\right].  
\label{eq:ex19}
\end{equation}
The three-- and four-- point contrast bounds on the effective conductivity are written by this parameter (see (21.33)--(21.35) and  (21.42)--(21.44) from \cite{Torq}). 

\begin{remark}
Formula \eqref{eq:ex19} is valid for arbitrary distributed circular inclusions. Any other shape of inclusions can be approximated by a disk packing. %that can extends applications of \eqref{eq:ex15} and \eqref{eq:ex19} to other shapes.    
\end{remark}

\begin{remark}
In accordance with \eqref{intEqS1} computation of the Torquato--Milton parameter $\zeta_1$ for general shapes of inclusions can be reduced to a sum of the triple integrals
\begin{equation}
%\sum_{k,k_1,k_2=1}^N 
\int_{L_k}\int_{L_{k_1}}\int_{L_{k_2}}  t_2 \overline{E_1(t_2-t_1)}E_1(t_1-t)  \textrm{d}\overline{t} \textrm{d}t_1 \textrm{d}\overline{t_2}
\label{eq:ex20}
\end{equation}    
that should be easier than computation of the corresponding triple integral (20.66) from \cite{Torq} on the 3--point correlation function which is computed by another triple integral. The integral \eqref{eq:ex20} can be calculated by residues for algebraic curves (see the case of ellipses discussed in \cite{Mit2009}).
\end{remark}

\section{Conclusion}\label{Discussion}

The iterative scheme \eqref{intEqS1} was constructively applied to special inclusions. An exact formula for regular array of disks (and solution of the problem \eqref{2.6psi} for $N=1$) was obtained in \cite{Mit1998}, \cite{Mit2007} (see survey \cite{MRP}). The term "exact formula" should be explained precisely. Here, the exact formula means that $\Lambda_e$ is written as an expression which contains the geometrical parameters (the fundamental translation vectors $\omega_j$ of the lattice $\mathcal Q$, the radius of the disk) and the physical parameter $\rho$ explicitly in symbolic form. The exact formula does not contains any parameter calculated by a numerical method (by an integral equation method or by truncation of an infinite system of equations). The exact formulae for $\Lambda_e$ in \cite{MRP} was written in the form of the series similar to \eqref{contr1} in which all the coefficients $C_n$ were exactly written.         

Approximate analytical formulae for $\Lambda_e$ were obtained for arbitrary non--overlapping locations of different disks in \cite{MitOpt}  and of ellipses in \cite{Mit2009}. Here, the term "approximate analytical formula" means that the effective conductivity $\lambda_e$ was written in the form of the series \eqref{contr1} in which the first few coefficients $C_n$ were exactly written. 
Other results obtained by the GMS are discussed also in Sec.\ref{Cluster2}.

In the previous works, the absolute convergence of the
method was proved under geometrical restrictions \cite{Mikhlin}.
Roughly speaking, these restrictions mean that each inclusion is far a
way from others (see \cite{Torq} and many papers cited therein). However, it was proved in \cite{MiR} that a modified method always uniformly converges, i.e., these geometrical restrictions are redundant. This is an interesting example of the difference between absolute and uniform convergence which shows that estimations on the absolute values or on the norm are too strong in comparison to the study of the uniform convergence \cite{Mit2011a}. 

The inclusions $D_k$ can have exotic shapes that complicates the criterion of the dilute composite. This difficulty can be overcame by a conformal mapping of $D$ onto a circular multiply connected domain. The capacity \cite{Henrici} (effective conductivity) is invariant under conformal mappings. Next, application of the functional equations for circular inclusions of different radii \cite{MitOpt} can be applied to estimate their interactions.

The GMS is effective in symbolic and numerical computations for non--overlapping inclusions of simple shapes (disks and ellipses) in the deterministic statement. When an approximate analytical formula is deduced, the ensemble average is calculated by the straight forward computations. It is interesting to combine numerical methods effective for one--inclusion problems with the analytically described interactions between inclusions by the GSM in Sec.\ref{SecGen}.

\section{Acknowledgements}
The author is grateful to Barbara Gambin and Ryszard Wojnar for the fruitful discussion on random composites during his visit to IPPT PAN (Warsaw).

\section*{Appendix. Convergence of the GMS}\label{SecApp}
This Appendix is devoted to study of the integral equations \eqref{eq:B1}, \eqref{intEqB} in the space $\mathcal A$. These equations can be shortly written in the form \eqref{intEqA}. We prove that equation \eqref{intEqA} has the unique solution for $|\rho|<1$ given by the series \eqref{eq:intEqA} obtained from \eqref{intEqA} by application of the successive approximations.  

We use the following general result from \cite{KrZab}. 
\begin{theorem} \label{th con} 
Let $B$ be a linear bounded operator in a Banach space $\mathcal{B}$. 
If for any element $f\in \mathcal{B}$ and for any complex number $\nu$ 
satisfying the inequality $|\nu| \leq 1$ equation
\begin{equation}
\label{eq:x1}
\Psi= \nu B \Psi+f
\end{equation}
has a unique solution, then the unique solution of the equation 
\begin{equation}
\label{eq:x2}
\Psi= B \Psi +f
\end{equation}
can be found by the method of successive approximations. The approximations converge in $\mathcal{B}$ to the solution
\begin{equation}
\label{eq:x3}
\Psi= \sum_{k=0}^{\infty} B^k f.
\end{equation}
\end{theorem}

Theorem \ref{th con} can be applied to equation $\Psi= \rho A \Psi +1$ (see \eqref{intEqA}) in the Banach space $\mathcal A$.

\begin{theorem} \label{th int} 
Let $|\rho| <1$. Equation $\Psi= \rho A \Psi +1$ has a unique solution. This solution can be found by the method of successive approximations convergent in the space $\mathcal A$.
\end{theorem} 

Proof. Let $|\nu| \leq 1$. Consider equations 
\begin{equation}
\begin{array}{rr}
\psi_k(z) = -\nu \rho \sum_{m=1}^N \frac{1}{2\pi i} \int_{L_m} \overline{n_m^2(t)}\;\overline{\psi_m(t)}E_1(t-z) \textrm{d}t+f_k(z),
\;z \in D_k
\\
(k=1,2,\ldots, N),
\\
\\
\psi_k(t) = -\nu \rho  \left[\frac{1}{2}\overline{n_k^2(t)}\;\overline{\psi_k(t)}+ \sum_{m=1}^N \frac{1}{2\pi i} \int_{L_m} \overline{n_m^2(t)}\;\overline{\psi_m(t)}E_1(t-z) \textrm{d}t\right] +f_k(t),
\\
\;t \in L_k \;(k=1,2,\ldots, N),
\end{array}
 \label{eq:2U}
\end{equation}
where $f_k \in \mathcal A_k$. %Equations on $L_k$ has a form similar to  \eqref{intEqB}. 

Let $\psi_k (z)$ be a solution of \eqref{intEqB}. Introduce the function $\psi (z)$ analytic in $D$ and H\"{o}lder continuous in its closure as follows
\begin{equation}
\label{eq:2V} 
\psi (z)=-\nu \rho \sum_{m=1}^n \frac{1}{2\pi i} \int_{L_m} \overline{n_m^2(t)}\;\overline{\psi_m(t)}E_1(t-z) \textrm{d}t, \quad z \in D.
\end{equation}
One can see that $\psi (z)$ is doubly periodic.
The expression in the right hand part of (\ref{eq:2V}) can be considered as Cauchy's integral on torus 
\begin{equation}
\label{eq:2C}
\Phi(z) = \frac{1}{2\pi i} \int_{L} \mu(t)E_1(t-z) \textrm{d}t 
\end{equation} 
for $z \in Q\backslash L$ with the density $\mu(t) =-\nu \rho \overline{n_m^2(t)}\;\overline{\psi_m(t)}$ on $L_m$. Using the property \eqref{jump} of \eqref{eq:2C} and (\ref{eq:2U}) on $L_k$ we arrive at the 
$\mathbb{R}$--linear conjugation relation on each fixed curve $L_k$
\begin{equation}
\label{eq:2W} 
\psi_k(t)-f_k(t)-\psi(t) =-\nu \rho \overline{n_k^2(t)}\;\overline{\psi_k(t)}, \quad t \in L_k.
\end{equation} 
Here, $\Phi^+(t) = \psi_k (t)-f_k(t)$ and $\Phi^-(t) =\psi(t)$. Integration of  \eqref{eq:2W} along $L_k$ yields 
\begin{equation}
\label{eq:3a} 
\phi(t) = \phi_k(t)-\nu \rho \;\overline{\phi_k(t)}-g_k(t)+\gamma_k, \quad t \in L_k\; \;(k=1,2,\ldots, N),
\end{equation} 
where the functions $\phi(z)$, $\phi_k(z)$ are analytic in $D$, $D_k$ and belong to $\mathcal C^{(1,\alpha)}$ in the closures of the domains considered; $g_k(z)$ is a primitive function of $f_k(z)$; $\gamma_k$ is an arbitrary constant. Moreover, the function $\phi(z)$ is quasi--periodic, hence 
\begin{equation}
\label{eq:3c}
\phi(z)=\varphi(z)+cz,
\end{equation} 
where $\varphi(z)$ is periodic and $c$ is a constant. The constant $\gamma_k$ can be included into $\varphi_{k}(z)$ by the representation $\gamma_k = \alpha_k - \nu \rho \;\overline{\alpha_k}$. Then, \eqref{eq:3a} becomes
\begin{equation}
\label{eq:3b} 
\varphi(t) = \varphi_{k}(t)-\nu \rho \;\overline{\varphi_{k}(t)}-g_k(t)-ct, \quad t \in L_k,\;
(k=1,2,\ldots, N),
\end{equation} 
It follows from \cite{BM2013} and Appendix B of \cite{Mit1999} that the general solution of the problem \eqref{eq:3b} has the following structure
\begin{equation}
\label{eq:3e} 
\varphi(z) = \varphi^{(1)}(z)+\varphi^{(2)}(z), \quad 
\varphi_{k}(z) = \varphi_{k}^{(1)}(z)+\varphi_{k}^{(2)}(z),
\end{equation} 
where 
\begin{equation}
\label{eq:3f} 
\varphi^{(1)}(t) = \varphi^{(1)}_{k}(t)-\nu \rho \;\overline{\varphi^{(1)}_{k}(t)}-g_k(t), \quad t \in L_k,\;
(k=1,2,\ldots, N)
\end{equation} 
and
\begin{equation}
\label{eq:3g} 
\varphi^{(2)}(t) = \varphi^{(2)}_{k}(t)-\nu \rho \;\overline{\varphi^{(2)}_{k}(t)}-c t, \quad t \in L_k,\;
(k=1,2,\ldots, N).
\end{equation} 
Each of the problems \eqref{eq:3f} and \eqref{eq:3g} (with fixed $c$) has a unique solution. The unique solution of the problem \eqref{eq:3g} can be easily found as $\varphi^{(2)}(z)=c z$ and $\varphi_k^{(2)}(z)=0$. 
Then, differentiation of \eqref{eq:3e}--\eqref{eq:3g} implies that the general solution of the problem \eqref{eq:2W} has the form
\begin{equation}
\label{eq:3h} 
\psi(z) = \psi^{(1)}(z)+c,
\end{equation}
where the function $\psi_{k}^{(1)}(z)$ is uniquely determined from the problem
\begin{equation}
\label{eq:4a} 
\psi^{(1)}(t) =\psi^{(1)}_k(t)+\nu \rho \overline{n_k^2(t)}\;\overline{\psi^{(1)}_k(t)}-f_k(t), \; t \in L_k\;
(k=1,2,\ldots, N).
\end{equation} 

Comparison of \eqref{eq:2W} and \eqref{eq:4a} proves the theorem.

The series 
\begin{equation}
\label{eq:Psi-a} 
\Psi = \sum_{k=0}^{\infty} \rho^kA^k1
\end{equation} 
converges absolutely for $|\rho|<1$, since it can be represented in the form $\Psi = \sum_{k=0}^{\infty} s^k \rho_0^k A^k1$ with $s \rho_0 =\rho$ and $|\rho|<|\rho_0|<1$. The latter series converges absolutely with the rate $|s|$ because the series $\sum_{k=0}^{\infty} \rho_0^k A^k1$ converges uniformly; hence, its general term tends to zero as $k \to \infty$.

Uniform convergence of \eqref{eq:Psi-a} for $|\rho|=1$ can be justified by use of the arguments form \cite{MiR}.


\begin{thebibliography}{999}

\bibitem{Adler}
Adler PM, Thovert J-F, Mourzenko VV (2012) Fractured porous media. Oxford University Press, Oxford

\bibitem{akh}
Akhiezer NI (1990) Elements of the Theory of Elliptic Functions. RI: American Mathematical Society, Providence
%
%\bibitem{andr}
%{\sc I.V. Andrianov,  V.V. Danishevskyy and D. Weichert},  {\em Simple estimations on effective transport properties of a random composite material with cylindrical fibres}, ZAMP, 2 (2008), pp.~889--903.

\bibitem{Chib}
Chibrikova LI (1977) Fundamental boundary value problems for the analytic functions. Kazan Univ. Publ., Kazan (in Russian)

\bibitem{andr1}
Bolshakov VI,  Andrianov IV,  Danishevskyy VV (2008) Asymptotic Methods for Calculation of Composite Materials with Microstructure. Porogi, Dnipropetrovs'k (in Russian)

\bibitem{BKN}
Berlyand L, Kolpakov AG, Novikov A.  (2012) Introduction to the Network Approximation Method for Materials Modeling. Cambridge University Press, Cambridge 

\bibitem{BiM2001}
Berlyand L, Mityushev V (2001) Generalized Clausius-Mossotti formula for random composite with circular fibers. J Statist. Phys. 102: 115-145 

\bibitem{BiM2005}
Berlyand L, Mityushev V (2005) Increase and decrease of the effective conductivity of a two phase composites due to polydispersity. J Stat. Phys. 118: 481-509

\bibitem{BM2013}
Bojarski B, Mityushev V (2013) R-linear problem for multiply connected domains and alternating method of Schwarz. J. Math. Sci. 189: 68-77

\bibitem{CzNM2012}	
Czapla R, Nawalaniec W, Mityushev V (2012) Effective conductivity of random two-dimensional composites with circular non-overlapping inclusions. Comput. Mat. Sci. 63: 118-126 

\bibitem{CzNM2012-2}
Czapla R, Nawalaniec W, Mityushev V  (2012) Simulation of representative volume elements for random 2D composites with circular non-overlapping inclusions. Theoretical and Applied Informatics 24: 227-242

\bibitem{Fil1970}
Grigolyuk EI, Filshtinsky LA (1970) Perforated Plates and Shells. Moscow, Nauka  (In Russian)

\bibitem{Fil1991}
Grigolyuk EI, Filshtinsky LA (1991) Periodical Piece--Homogeneous Elastic Structures. Moscow, Nauka  (In Russian)

\bibitem{Fil1994}
Grigolyuk EI, Filshtinsky LA (1994)
Regular Piece-Homogeneous Structures with defects.  Moscow, Fiziko-Matematicheskaja Literatura (In Russian)

\bibitem{Henrici}
Henrici P (1986) Applied and computational complex analysis, v. 3, Wiley, New York etc 

\bibitem{andr2}
Kalamkarov AL,  Andrianov IV,  Danishevskyy VV (2009) Asymptotic homogenization of composite materials and structures.  Appl. Mech. Rev., 62:030802-1--030802-20. 

\bibitem{Kanit}
Kanit T, Forest S, Galliet I, Mounoury V, Jeulin D (2003) Determination of the size of the representative volume element for random composites: statistical and numerical approach. Int. J. Solids and Structures 40: 3647-3679

\bibitem{Kolpakov}
Kolpakov AA,  Kolpakov AG (2009) Capacity and Transport in Contrast Composite Structures: Asymptotic Analysis and Applications. CRC Press Inc., Boca Raton etc

\bibitem{KrZab}  
Krasnosel'skii MA et al. (1972)  Approximate Methods for Solution of Operator Equations. Wolters-Noordhoff Publ., Groningen

\bibitem{Kress}
Kress R (1999) Linear Integral Equations. Springer-Verlag, New York, Berlin Heideberg

\bibitem{LL}
Landau LD, Lifshitz EM, Pitaevskii LP (1982) Electrodynamics of Continuous Media. Nauka, Moscow (in Russian)
%
%\bibitem{McPh}
%{\sc  W. T.Perrins, D.R. McKenzie and R.C. McPhedran}, {\em Transport Properties of Regular Arrays of Cylinders}, Proc. R. Soc. Lond. A369, (1979), pp.~207--225.
%
%\bibitem{McPh2}
%{\sc R.C. McPhedran},
%{\em Transport Properties of Cylinder Pairs and of the Square Array of Cylinders}, Proc. R. Soc. Lond. A408, (1986), pp.~31--43.
%
%\bibitem{McPh3}
%{\sc R.C. McPhedran and G. W. Milton},
%{\em Transport Properties of Touching Cylinder Pairs and of the Square Array of Touching Cylinders}, Proc. R. Soc. Lond. A411, (1987), pp.~313--326.
%
%\bibitem{McPh4}
%{\sc R.C. McPhedran, L. Poladian and G. W. Milton},
%{\em Asymptotic Studies of Closely Spaced, Highly Conducting Cylinders}, Proc. R. Soc. Lond. A415, (1988), pp.~185--196.

\bibitem{Mikhlin}
Mikhlin SG (1964) Integral Equations and Their Applications
to Certain Problems in Mechanics, Mathematical Physics and
Technology. 2nd rev. ed., Macmillan, NY

\bibitem{Milton}
Milton GW  (2002) The theory of composites, Cambridge University Press, Cambridge 

\bibitem{Mit1993}
Mityushev VV  (1993) Plane problem for the steady heat conduction of material with circular inclusions. Arch. Mech. 45: 211-215

\bibitem{Mit1994}
Mityushev V (1994) Solution of the Hilbert boundary value problem for a multiply connected domain. Slupskie Prace Mat-Przyr. 9a: 33-67 

\bibitem{Mit1995}
Mityushev V (1995) Generalized method of Schwarz and addition theorems in mechanics of materials containing cavities. Arch. Mech. 47: 1169-1181 
%
%\bibitem{Mit3D}
% V. Mityushev,  N. Rylko,
%Boundary value problems, the Poincare series, the method of Schwarz and composite materials. Int. Congres IMACS 97, Berlin, v.1 (1997) 165--170.

\bibitem{Mit1998}
Mityushev V (1998) Steady heat conduction of the material with an array of cylindrical holes in the non-linear case. IMA J Appl. Math. 61: 91-102
%
%\bibitem{MitArch}
% V. Mityushev,  
%Thermoelastic plane problem for material with circular inclusions, {\em Arch. Mech.} {\bf 52} (2000) 915--932.

\bibitem{Mit1999}	
Mityushev V (1999)
Transport properties of two-dimensional composite materials with circular inclusions. Proc. R. Soc. Lond. A 455: 2513-2528

\bibitem{Mit2005}
Mityushev V (2005) $\mathbb R$--linear problem on torus and its application to composites. Complex Variables 50: 621-630

\bibitem{Mit2007}
Mityushev V (2007) Exact solution of the R-linear problem for a disk in a class of doubly periodic functions. J. Appl. Funct. Anal. 2: 115-127 

\bibitem{Mit2006}
Mityushev V (2006) Representative Cell in Mechanics of Composites and Generalized Eisenstein-Rayleigh Sums. Complex Variables and Elliptic Equations 51: 1033-1045

\bibitem{Mit2009}
Mityushev V (2009) Conductivity of a two-dimensional composite containing elliptical inclusions. Proc. Roy. Soc. Lond. A 465: 2991-3010

\bibitem{Mit2011a}
Mityushev V (2011) Riemann-Hilbert problems for multiply connected domains and circular slit maps. Computational Methods and Function Theory  11: 575-590

\bibitem{Mit2012}
Mityushev V, Rylko N (2012) Optimal distribution of the non-overlapping conducting disks. Multiscale Model. Simul. 10: 180-190
%
%\bibitem{Mit2013a}
%Mityushev V., Rylko N., A fast algorithm for computing the flux around non--overlapping disks on the plane, Mathematical and Computer Modelling 57 (2013) 1350--1359.

\bibitem{Mit2013b}
Mityushev V, Rylko N (2013) Maxwell's approach to effective conductivity and its limitations, The Quarterly Journal of Mechanics and Applied Mathematics; doi: 10.1093/qjmam/hbt003

\bibitem{MiR}  
Mityushev VV, Rogosin SV (2000) Constructive methods for linear and non-linear boundary value problems of the analytic function. Theory and applications. Chapman \& Hall / CRC, Boca Raton etc.

\bibitem{MitOpt}	
Mityushev V (2001) Transport properties of doubly periodic arrays of circular cylinders and optimal design problems. Appl. Math. Optimization 44: 17-31

\bibitem{MRP}
Mityushev VV, Pesetskaya E, Rogosin SV (2008) Analytical Methods for Heat Conduction in Composites and Porous Media, in
Cellular and Porous Materials: Thermal Properties Simulation and Prediction, A. \"{O}chsner, G. E. Murch, M. J. S. de Lemos, eds., Wiley, 121-164

%\bibitem{Mit2011}
%{\sc V. V. Mityushev},  {\em Riemann--Hilbert problems for multiply connected domains and circular slit maps}, Comp. Meth. Funct. Theory, 11, (2011), pp.~575--590.
%
\bibitem{Movchan}  
Movchan AB, Movchan NV, Poulton CG (2002) Asymptotic Models of Fields in Dilute and Densely Packed Composites, Imperial College Press, London 

\bibitem{Ray}  
Rayleigh Lord (1892) On the influence of obstacles arranged in rectangular order upon the properties of medium. Phil Mag
34: 481-502

\bibitem{SBG}  
Smith B, Bj\"{o}rstad P, Gropp W (1996) Domain decomposition. Parallel multilevel methods for elliptic partial differential
equations. Cambridge University Press, Cambridge

\bibitem{Telega}
Telega JJ (2004) Stochastic homogenization: convexity and nonconvexity,  eds. P.P. Casta\~{n}eda, J.J. Telega, B. Gambin, Nonlinear Homogenization and its Applications to Composites, Polycrystals and Smart Materials, NATO Science Series,  Kluwer Academic Publishers, Dordrecht, 305-346

%\bibitem{Ting}
%Ting, T.C.T.: Anisotropic Elasticity. Theory and Applications. Oxford University (1996)

\bibitem{Torq}
Torquato S ( 2002) Random Heterogeneous Materials: Microstructure and Macroscopic Properties. Springer-Verlag, NY

\bibitem{weil}
Weil A (1999) Elliptic Functions according to Eisenstein and Kronecker, Springer-Verlag, Berlin etc 

\bibitem{Szcz}
Szczepkowski J, Malevich AE, Mityushev V (2003) Macroscopic properties of similar arrays of cylinders. Quart J Appl Math Mech 56: 617-628 

\end{thebibliography}
\end{document}